# A Utility Framework for Bounded-Loss Market Makers

**Yiling Chen**
Yahoo! Research
45 W. 18th St. 6th Floor
New York, NY

**David M. Pennock**
Yahoo! Research
45 W. 18th St. 6th Floor
New York, NY

## Abstract

We introduce a class of *utility-based market makers* that always accept orders at their risk-neutral prices. We derive necessary and sufficient conditions for such market makers to have bounded loss. We prove that hyperbolic absolute risk aversion utility market makers are equivalent to weighted pseudospherical scoring rule market makers. In particular, Hanson's logarithmic scoring rule market maker corresponds to a negative exponential utility market maker in our framework. We describe a third equivalent formulation based on maintaining a cost function that seems most natural for implementation purposes, and we illustrate how to translate among the three equivalent formulations. We examine the tradeoff between the market's liquidity and the market maker's worst-case loss. For a fixed bound on worst-case loss, some market makers exhibit greater liquidity near uniform prices and some exhibit greater liquidity near extreme prices, but no market maker can exhibit uniformly greater liquidity in all regimes. For a fixed minimum liquidity level, we give the lower bound of market maker's worst-case loss under some regularity conditions.

## 1 Introduction

A financial market is a central place where people converge to trade commodities or securities of uncertain value. In a well-functioning market, the resulting prices can reflect a wealth of information about the expected value of the security—in ideal circumstances the sum total of all information available to all traders. Most financial markets are designed to satisfy demand for trade, and price discovery is a beneficial side effect.

We are interested in the design of *prediction markets*,[1] or markets where price discovery is the designer's end goal and trading is a means to that end. For example, suppose a policymaker seeks a forecast of the likelihood of an avian flu outbreak in 2008. He may float a security paying $1 if and only if an outbreak actually occurs in 2008, hoping to attract traders willing to speculate on the outcome. With sufficient liquidity, traders will converge to a consensus price reflecting their collective information about the value of the security, which in this case corresponds to the market's estimate of the probability of outbreak. Empirically, prediction markets often yield better forecasts than other methods across a diverse array of settings including politics [Forsythe et al., 1999, Berg et al., 2001], sports [Debnath et al., 2003], and business [Chen and Plott, 2002].

When a market fails to attract sufficient traders it may break down. In fact, many financial and betting markets are thinly traded, resulting in little or no price discovery. In prediction markets, the problem can be expected to be even worse, since meeting trader demand is not the market's primary design goal. *Combinatorial prediction markets* [Fortnow et al., 2004, Hanson, 2003, 2007, Chen et al., 2007] further exacerbate the problem by dividing traders' attention among an exponential number of outcomes of a joint random variable, making the likelihood of finding agreeable bilateral trades even more remote. Thin markets lead to a "chicken and egg" problem where few traders care to participate because other traders are scarce, potentially spiraling the market into failure. In fact, beyond mere apathy, traders have an active incentive *not* to post an order in a thin market, since posting an order reveals information with little chance of any benefit; this relates to the so-called *no-trade theorems* [Milgrom and Stokey, 1982] for speculative markets.

---

[1]Prediction markets are also referred to as information markets, (Arrow-Debreu) securities, contingent claims or contracts, event markets or futures, and idea futures.

An *automated market maker* can improve the liquidity of a prediction market. The market maker continually announces prices offering both to buy and to sell some quantity of the security, adjusting his prices in programmatic response to trader demand. With the addition of a market maker, a prediction market can incorporate the information of even a solitary trader, and can aggregate demand from traders who arrive and depart at non-overlapping times. If the market maker can be expected to lose money on average, then the trading game becomes positive sum, circumventing the no-trade theorems.

Traditionally, market makers are human decision makers seeking to earn a profit. In contrast, a prediction market designer may well *subsidize* a market maker that expects to lose some money, in return for improving trader incentives, liquidity, and price discovery. The market maker's loss can be seen as the cost of gathering information for more accurate forecasts. Of course, the market operator cannot afford to lose arbitrary amounts of money; typically he or she will want to set a maximum bound on the market maker's loss, ensuring that no matter what happens, the loss cannot exceed the bound.

In Section 3, we develop a class of bounded-loss automated market makers that we call *utility-based market makers*. Each market maker has some utility function for money. However, instead of attempting to maximize his expected utility, the market maker seeks to keep his expected utility constant at all times, ensuring that his utility will never decrease. Equivalently, the market maker always accepts (infinitesimal quantity) buy and sell orders at his risk-neutral probabilities; that is, the market maker's risk-neutral probabilities become the (instantaneous) market prices. As the market maker accepts orders, his risk-neutral probabilities (and thus the prices) change. We prove necessary and sufficient conditions on the utility function for the market maker's loss to be bounded.

We describe how our utility-based framework fits within other market maker frameworks. In Section 4, we show that a utility-based market maker with (nonzero) hyperbolic absolute risk aversion (HARA) is equivalent to a weighted pseudospherical *market scoring rule market maker* [Hanson, 2003, 2007]. That is, for any non-linear utility function in the class of HARA utility functions, there is an equivalent scoring function in the class of weighted pseudospherical scoring functions, such that the implied market makers are behaviorally equivalent. In particular, our negative exponential utility market maker corresponds to Hanson's logarithmic scoring rule market maker [Hanson, 2003, 2007]. Our logarithmic utility market maker—which has never been proposed before to our knowledge and corresponds to an atypical scoring rule—has some advantages (and disadvantages) when compared to Hanson's logarithmic scoring rule market maker. In Section 5, we show how both ours and Hanson's frameworks can be implemented using a cost function methodology. The cost function records the total amount spent by traders as a function of the total quantities of shares outstanding. We argue that, when a cost function can be derived, it offers the most intuitive and straightforward method for implementing these market makers in practice. Unfortunately, in some cases we cannot derive explicit cost functions and must resort to indirect or implicit numerical implementations.

In Section 6, we examine the trade-off between a market maker's worst-case loss and the *instantaneous liquidity*, the continuous-price analog of the liquidity, under some regularity conditions. For a fixed bound of worst-case loss, we show that no market maker can uniformly exhibit higher instantaneous liquidity than another in all price regimes. We prove the lower bound of the market maker's worse-case loss for a given minimum instantaneous liquidity level. We conclude our study in section 7. Due to limit of space, we omit proofs of theorems, which can be obtained as an Appendix by request.

## 2 Background

Let $v$ represent a discrete or discretized random variable to be predicted, with $N$ mutually exclusive and exhaustive outcomes. $\vec{r} = (r_1, r_2, ..., r_N)$ be a probability estimate for the random variable $v$. A *scoring rule* is a sequence of scoring functions, $S = s_1(\vec{r})$, $s_2(\vec{r})$, ..., $s_N(\vec{r})$, such that a score $s_i(\vec{r})$ is assigned to $\vec{r}$ if outcome $i$ of the random variable $v$ is realized. A *proper scoring rule* [Winkler, 1969] is a scoring rule that motivates truthful reporting.

Hanson [2003, 2007] proposes a mechanism where a patron subsidizes an automated market maker in order to improve liquidity and overcome no-trade reticence. The patron is guaranteed up front not to lose more than a fixed constant subsidy regardless of how many trades are processed or what outcome eventually occurs. Hanson's mechanism is called a *market scoring rule market maker* (MSR), so-named because the mechanism can be thought of as a market version of a proper scoring rule. Conceptually, the market maker with a proper scoring rule $S$ begins by setting an initial probability estimate, $\vec{r}^0$. Every trader can change the current probability estimate to a new estimate of his choice as long as he agrees to pay the market maker the scoring rule payment associated with the current probability estimate and receive the scoring rule payment associated with the new estimate. Myopically, this modified scoring rule still incents the trader to reveal his true probability estimate.

Because traders change the probability estimate in sequence, the MSR market maker in fact only pays the last trader and receives payment from the first trader. The market maker incurs the maximum loss when the final probability estimate assigns probability 1 to the true outcome, i.e. $s_j(\vec{e_j}) - s_j(\vec{r}^0)$, where $j$ is the index for the true outcome and $e_j$ is the vector whose $j$th element is 1 and all other elements are 0. If the market maker starts with a uniform distribution, the maximum amount to lose is bounded by $b \log N$ when using a *logarithmic scoring rule*,

$$s_i(\vec{r}) = b \log(r_i) \qquad (b > 0), \qquad (1)$$

and by $\frac{(N-1)b}{N}$ when using a *quadratic scoring rule*,

$$s_i(\vec{r}) = 2br_i - b \sum_j r_j^2 \qquad (b > 0). \qquad (2)$$

## 3 Utility-Based Market Makers

We propose a new class of market makers who have a utility function and set prices equal to their risk-neutral probabilities. We show some general conditions under which the utility-based market maker has provably bounded loss.

### 3.1 Risk-Neutral Probability

Let an agent have a state-independent utility function of money, $u(m)$, and a subjective probability estimate, $\vec{\pi}$, for the forecast variable. Let $\vec{m}$ represents his wealth vector across outcome states. Then, the agent's risk-neutral probabilities [Jackwerth, 2000] are the normalized products of his subjective probabilities and marginal utilities,

$$p_i = \frac{\pi_i u'(m_i)}{\sum_j \pi_j u'(m_j)} \qquad \forall i. \qquad (3)$$

In a securities market setting, risk-neutral probabilities are the price levels that the agent is indifferent between buying or selling an infinitely small number of shares. The subjective probability estimate and risk-neutral probability estimate are identical when the agent is indifferent to risk. However, for a risk-averse agent they differ by a risk aversion adjustment.

### 3.2 Utility-Based Market Makers

A *utility-based market maker* has a utility function for money $u(m)$ and a subjective probability estimate $\vec{\pi}$. The market maker myopically equates the instantaneous security prices to his current risk-neutral probabilities (3), and is always willing to accept infinitely small buy or sell orders at these prices.

The market maker's wealth across all outcome states changes when he accepts orders in the market. Suppose a trader buys an infinitely small number of security $i$, denoted $\epsilon_i$. Then the market maker's wealth in state $j \neq i$ increases by $p_i \epsilon_i$, while his wealth in state $i$ decreases by $(1-p_i)\epsilon_i$, because the market maker pays off \$1 per share in state $i$. Let $\vec{q} = (q_1, q_2, ..., q_N)$ be the vector of the total quantities of outstanding shares. Then, $\partial m_j / \partial q_i = p_i$ when $j \neq i$, and $\partial m_i / \partial q_i = p_i - 1$. The following equation system thus defines a utility-based market maker:

$$\begin{cases} p_i = \dfrac{\pi_i u'(m_i)}{\sum_j \pi_j u'(m_j)} & \forall i \\ \dfrac{\partial m_j}{\partial q_i} = p_i - I_{ij} & \forall i, j \end{cases} \qquad (4)$$

where $I_{ij}$ is an indicator function that equals 1 when $i = j$ and 0 otherwise.

**Lemma 1.** *At any time of the market, a utility-based market maker who sets prices according to (4) satisfies*

$$\sum_j \pi_j u(m_j) = k, \qquad (5)$$

*where $k$ is a constant.*

Lemma 1 is implied by the concept of risk-neutral probability. It is proved by showing that the partial derivatives of both sides of equation (5) relative to $q_i$ are zeros for any $i$.

Thus, a utility-based market maker does not maximize his expected utility. Instead, the market maker starts the market with some initial expected utility and then keeps this expected utility level during the whole process of trading.

### 3.3 Loss of Market Makers

A market maker can choose any initial feasible wealth vector $\vec{m}^0$ to start the market and $k$ equals $\sum_j \pi_j u(m_j^0)$. Theorem 2 characterizes properties of the market maker's utility function that guarantee bounded loss.

**Theorem 2.** *For a real-valued, continuous and strictly increasing utility function $u(m)$, any fixed subjective probability estimate $\vec{\pi}$ whose elements are non-zeros, and a utility-based market maker who sets prices according to (4), the necessary and sufficient condition for the market maker to have bounded loss for any feasible expected utility level $k$ is that at least one of the following conditions is satisfied: (1) The domain of $u(m)$ is bounded below; (2) The range of $u(m)$ is bounded above but not bounded below.*

We prove Theorem 2 with the assistance from Lemma 1. A market maker has bounded loss if and only if there exists a real constant $l$ such that for any $\vec{m}$ that meets (5) in Lemma 1, $l \leq m_j$ is satisfied for all $j$. Theorem 2 indicates that neither linear functions nor strictly convex functions defined on $(-\infty, +\infty)$ guarantee bounded loss.

## 4 Relationship Between MSR and Utility-Based Market Makers

A trader with a probability estimate on the forecast variable $v$ can interact with either a MSR market maker or a utility-based market maker for profit or loss. Hence, a natural question to ask is how these two classes of market makers are related. In this section, we establish the equivalence relationship between them for a class of utility functions, the HARA utility class, and a class of proper scoring rules, the weighted pseudospherical scoring rules.

### 4.1 HARA Utility Class

The *hyperbolic absolute risk aversion (HARA)* class of utility functions contains most popular parametric families of utilities, including constant absolute risk aversion (CARA) family and constant relative risk aversion (CRRA) family. The generic form of a HARA utility function is

$$u(m) = \frac{1}{1-\gamma}\left(\gamma(M + \frac{\alpha}{\gamma}m)^{1-\gamma} - 1\right), \quad (6)$$

where $M$ is a real number, $\gamma$ is an extended real number, and $\alpha > 0$. The utility function is defined on the domain $M + \frac{\alpha}{\gamma}m \geq 0$, with strict inequality for $\gamma \geq 1$. Any affine transformation of (6) belongs to the HARA utility class and leaves all economic analysis unchanged. The characterization of the HARA utility class is that utility functions have linear absolute risk tolerance, $-\frac{u'(m)}{u''(m)} = \frac{M}{\alpha} + \frac{m}{\gamma}$, which is defined as the inverse of the absolute risk aversion coefficient [Arrow, 1970]. For this reason, the HARA utility class is also called the linear-risk-tolerance utility class.

Special choices of parameters $M$, $\gamma$, and $\alpha$ give rise to different (families of) utility functions. When $\gamma = 0$, linear or risk-neutral utility function $u(m) = \alpha m - 1$ is obtained. When $\gamma > 0$, (6) is the family of constant relative risk aversion (CRRA) utility functions. The limiting case where $\gamma \to 1$ gives the logarithmic utility function, $\lim_{\gamma \to 1} u(m) = \log(M + \alpha m)$. In the limit as $\gamma \to -\infty$ or $\gamma \to +\infty$, (6) becomes the negative exponential utility, $\lim_{\gamma \to \pm\infty} u(m) = -e^{-\alpha m}$, which is a constant absolute risk aversion (CARA) utility function. Utility functions corresponding to $\gamma < 0$ has domains that are bounded above, which do not satisfy the economics principle of nonsatiation. A more extensive review of the HARA utility class has been given by Feigenbaum [2003].

It can be verified that with the exception of the linear utility function, all HARA utility functions satisfy the condition in Theorem 2. Hence, a utility-based market maker using a non-linear HARA utility function is guaranteed to have bounded loss.

### 4.2 Weighted Pseudospherical Scoring Rules

Jose et al. [2006] in their recent work proposes a class of strictly proper scoring rules that generalizes the pseudospherical scoring rules, called *weighted pseudospherical scoring rules*. The scoring function of this class is

$$s_i(\vec{r}) = a_i + \frac{b}{\beta-1}\left[\left(\frac{\frac{r_i}{\pi_i}}{(\sum_j \pi_j(\frac{r_j}{\pi_j})^\beta)^{1/\beta}}\right)^{\beta-1} - 1\right], \quad (7)$$

where $b > 0$ and the reported probability estimate $\vec{r}$ is weighted by a base-line probability estimate $\vec{\pi}$. Scoring rule payments are based on the improvement of the reported probability estimate on the base-line probability estimate. If weighted by a uniform distribution, weighted pseudospherical scoring rules reduce to pseudospherical scoring rules. Weighted pseudospherical scoring rules are strictly proper for any real $\beta$. Choices of $\beta$ give rise to different scoring functions. For example, the limiting case where $\beta \to 1$ defines a weighted logarithmic scoring rule,

$$s_i(\vec{r}) = a_i + b\log(\frac{r_i}{\pi_i}). \quad (8)$$

It becomes the (unweighted) logarithmic scoring rule (1) when all $\pi_i$'s are equal. When $\beta = 2$ and $\vec{\pi}$ is uniform, (7) is the spherical scoring rule. The limit case as $\beta \to 0$ gives the following atypical scoring function,

$$s_i(\vec{r}) = a_i - b\frac{\pi_i}{r_i}\exp\left(\sum_j \pi_j \log(\frac{r_j}{\pi_j})\right). \quad (9)$$

The quadratic scoring rule (2), however, does not belong to this class. Because the payment when probability 1 is reported for the true outcome is finite for all weighted pseudospherical scoring rules, a MSR market maker's loss is bounded when using any one of them.

### 4.3 The Market Maker Equivalence Theorem

Jose et al. [2006] show that the weighted pseudospherical scoring rules can arise from the solution of an expected-utility-maximization problem of a forecaster against a betting opponent. A market with a market maker can be viewed as a sequence of two-person betting, each happening between a trader and the market maker. Taking this view, we show the equivalence between non-linear HARA-utility market makers and weighted pseudospherical market scoring rule market makers.

A bet between two agents, A and B, on the random variable $v$ can be interpreted as selecting a vector $\vec{m}$

such that the contingent wealth vectors for agents A and B are $\vec{m}$ and $-\vec{m}$ respectively and both agents are willing to accept it given their subjective probability estimates on $v$. Any trade in the market with a utility-based market maker is a betting between the market maker and the trader. Facing the existing wealth vector of the market maker, $\vec{m}^{old}$, the (risk-neutral) trader with probability estimate $\vec{r}$ wants to select a new wealth vector $\vec{m}^{new}$ for the market maker such that the trader's expected wealth, $\sum_i r_i[(-m_i^{new}) - (-m_i^{old})]$, is maximized. The market maker is willing to accept the wealth vector $\vec{m}^{new}$ only if it satisfies equation (5). Hence, any trade between the utility-based market maker and a trader with probability estimate $\vec{r}$ can be modeled by the following optimization problem,

$$\max_{\vec{m}} \quad -\sum_i r_i m_i \qquad (10)$$
$$\text{s.t.} \quad \sum_i \pi_i u(m_i) = k.$$

Based on (10), we obtain our *Market Maker Equivalence Theorem*.

**Theorem 3** (Market Maker Equivalence Theorem). *A utility-based market maker who has a subjective probability estimate $\vec{\pi}$ and a HARA utility function with $\gamma \neq 0$ is equivalent to a market scoring rule market maker who utilizes a $\vec{\pi}$-weighted pseudospherical scoring rule with $\beta = 1 - \frac{1}{\gamma}$.*

As expected, linear utility function ($\gamma = 0$), which does not guarantee a utility-based market maker's loss to be bounded according to Theorem 2, does not have a corresponding scoring rule. The following corollary arises directly by applying Theorem 3 with $\gamma = \pm\infty$ and $\gamma = 1$.

**Corollary 4.** *A negative exponential utility market maker is equivalent to a MSR market maker with a weighted logarithmic scoring rule (8). A logarithmic utility market maker is equivalent to a MSR market maker with the scoring rule defined in (9).*

If starting the market with uniform prior probabilities, a negative exponential utility market maker is equivalent to a logarithmic MSR market maker.

## 5 Cost-Function Formulation of Market Makers

### 5.1 Cost-Function Formulation

We describe an equivalent formulation of market makers based on maintaining a cost function that seems most natural for implementation purposes. The market contains a total of $N$ securities, each paying $1 per share if its corresponding outcome happens. $\vec{q}$ is the vector of all quantities of shares held by traders.

Then, the market maker works as follow: (1) The market maker utilizes a *cost function* $C(\vec{q})$ that records the total amount of money traders have spent as a function of the total number of shares held of each security. The market maker initiates the market with a quantity vector $\vec{q}^0$. (2) A trader who buys or sells any security or any bundles of securities in the market changes the the total number of outstanding securities, i.e. $\vec{q}$, from $\vec{q}^{old}$ to $\vec{q}^{new}$. The market maker charges the agent $C(\vec{q}^{new}) - C(\vec{q}^{old})$ dollars for the transaction. Negative quantities encode sell orders and negative "payments" represent sale proceeds earned by the trader. (3) At any time of the market, the going price of security $i$, $p_i(\vec{q})$, equals $\partial C/\partial q_i$. The price is the cost per share for purchasing an infinitesimal quantity of security $i$. The full cost for purchasing any finite quantity is the integral of price evaluated from $\vec{q}^{old}$ to $\vec{q}^{new}$, which equals $C(\vec{q}^{new}) - C(\vec{q}^{old})$. (4) Once the true outcome becomes known, the market maker pays $1 per share to traders holding the winning security.

Prices of all securities are restricted to lie in [0, 1] at all times, and the sum of prices always equals 1 to ensure no arbitrage. The cost and price functions thus satisfy the following property.

**Property 5.** *For a market maker with a cost function $C(\vec{q})$ and price functions $p_i(\vec{q})$ in an arbitrage-free market, the following equations hold,*

$$C(\vec{q} + a\vec{1}) = a + C(\vec{q}), \qquad (11)$$

*and*

$$p_i(\vec{q} + a\vec{1}) = p_i(\vec{q}) \quad \forall i, \qquad (12)$$

*where $a$ is any real constant.*

### 5.2 Cost Functions and Utility-Based Market Makers

We show that any utility-based market maker can be conveniently translated to a cost-function formulation.

**Theorem 6.** *A utility-based market maker has a cost function that is defined by*

$$\sum_j \pi_j u(C - q_j) = k, \qquad (13)$$

*where $k$ is a constant.*

Theorem 6 follows from Lemma 1, since $C(\vec{q})$ is the total money collected by the market maker and $q_j$ is the total money that the market maker needs to pay if outcome $j$ happens.

According to the *implicit function theorem* in mathematical analysis, we can see that if the utility function $u(m)$ is continuous, differentiable, and strictly increasing, there exists a unique function $C(\vec{q})$. If $C(\vec{q})$ does not have an explicit form, numerical methods can be used to calculate the value of $C$ for any given $\vec{q}$. Prices are the partial derivatives of the cost function.

Logarithmic utility and negative exponential utility are two widely used utility functions that both belong to the HARA utility class. The cost function corresponding to the logarithmic utility function, $u(m) = \log(b+m)$ with $b > 0$, is the implicit function defined by equation (13). If the event only has two outcomes and the market maker's subjective probability estimate is uniform, the explicit cost function for the market maker is

$$C(\vec{q}) = -b + \frac{1}{2}(q_1 + q_2) + \frac{1}{2}\sqrt{4b^2 + (q_1 - q_2)^2}. \quad (14)$$

For the negative exponential utility function, $u(m) = -e^{-\alpha m}$ with $\alpha > 0$, the market maker's cost function is

$$C(\vec{q}) = \frac{1}{\alpha}\log(\sum_j \pi_j e^{\alpha q_j}). \quad (15)$$

We omit the price functions here, which can be easily obtained by differentiation.

### 5.3 Cost Functions and Market Scoring Rules

Given the Market Maker Equivalence Theorem and Theorem 6, a weighted pseudospherical scoring rule market maker can be translated into the corresponding cost function formulation via his equivalent utility-based market maker. For scoring rules, such as quadratic scoring rule, that do not belong to the weighted pseudospherical scoring rule class, we describe how to make the translation directly.

Suppose the current probability estimate in MSR is $\vec{p}$. A trader who has probability estimate $\vec{p}'$ changes the market probabilities from $\vec{p}$ to $\vec{p}'$ and gets profit $s_i(\vec{p}') - s_i(\vec{p})$. Now suppose, in a market maker mechanism that offers $N$ mutually exclusive and exhausted securities, the current quantity vector is $\vec{q}$ and price vector is $\vec{p}$. A trader with probability estimate $\vec{p}'$ is myopically incented to buy or sell securities until the market price becomes $\vec{p}'$. His trading behavior changes the quantity vector to $\vec{q}'$. His profit when outcome $i$ happens is $(q_i' - q_i) - (C(\vec{q}') - C(\vec{q}))$. If the two mechanisms are equivalent, the trader should obtain the same profit no matter which outcome of the random variable is realized. Thus, without lose of generality, the following equation system establishes the equivalence between a MSR market maker and a cost-function formulation

$$\begin{cases} s_i(\vec{p}) = q_i - C & \forall i \\ \sum_i p_i = 1 \\ p_i = \frac{\partial C}{\partial q_i}. \end{cases} \quad (16)$$

Solving (16), we get the cost function for the MSR market maker with logarithmic scoring rule as in (1),

$$C(\vec{q}) = b\log(\sum_j e^{q_j/b}), \quad (17)$$

which is equivalent to the cost function (15) derived for negative exponential utility market maker by setting $\vec{\pi}$ to be uniform and with some variable substitution, verifying the stated equivalence result in Corollary 4. The equivalent cost function for a MSR market maker with a quadratic scoring rule (2) is

$$C(\vec{q}) = \frac{\sum_j q_j}{N} + \frac{\sum_j q_j^2}{4b} - \frac{(\sum_j q_j)^2}{4Nb} - \frac{b}{N}, \quad (18)$$

Price functions can be obtained by differentiation.

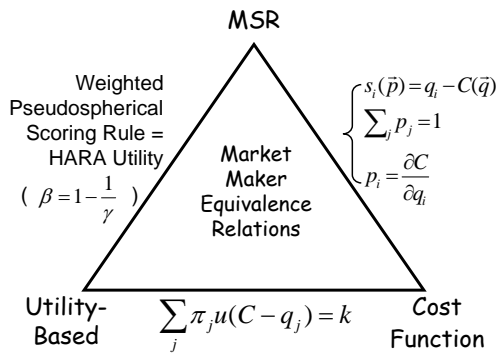

Figure 1: Market Maker Equivalence Relations

At this point we have established equivalence translations for restricted classes among utility-based market makers, market scoring rule market makers, and a third formulation of market makers based on maintaining a cost function. Figure 1 summarizes our equivalence results. The equivalence relations enable easy navigation among different formulations. The cost function formulation seems most natural for implementation purposes. Market scoring rules make many analysis such as loss of market makers straightforward, however directly converting them to cost-function formulations is not always easy. Utility-based market makers, whose cost functions are easy to find, connect the MSR with cost-formulation for a large class of scoring rules.

## 6 Liquidity and Market Maker Loss

### 6.1 Instantaneous Liquidity

In financial markets, high liquidity is often characterized by small *bid-ask spread*, which is the difference between the (ask) price at which a market maker is willing to accept buy orders and the (bid) price at which the market maker is willing to accept sell orders. It is common that any bid and ask prices set by the market maker are only valid for trading a fixed number of shares, after which the market maker adjusts prices and/or the bid-ask spread. Such a market maker essentially has a step-wise price function. Another concept that is often associated with liquidity is *market depth*. A market is considered deep if moving the price up (down) by a unit requires buying (selling) a large number of shares. A deep market often implies high liquidity.

For our continuous and differentiable price functions, these concepts of bid-ask spread, market depth, and hence liquidity for financial markets are not directly applicable. We define *instantaneous liquidity* to reflect the liquidity in such markets.

**Definition 1.** *The instantaneous liquidity for security $i$ at $\vec{q}$ is $\rho_i(\vec{q}) = \frac{1}{\partial p_i(\vec{q})/\partial q_i}$.*

The slope of the price function, $\partial p_i(\vec{q})/\partial q_i$, approximates the bid-ask spread for the differentiable price function $p_i(\vec{q})$. Although only one price rather than a bid and an ask is announced and the bid-ask spread is technically zero for infinitely small trades, the payment required for buying certain number of shares is typically higher than the proceeds received for selling the same number of shares at any time of the market. $\partial p_i(\vec{q})/\partial q_i$ approximates the price difference between buying one share of security $i$ from the market maker and selling one share to the market maker at the current state $\vec{q}$. Moreover, $\partial p_i(\vec{q})/\partial q_i$ inversely relates to the instantaneous depth of the market for security $i$. A smaller $\partial p_i(\vec{q})/\partial q_i$ means that it requires more shares to drive price up or down by one unit, hence implying a deeper market. Thus, the definition of instantaneous liquidity captures both the negative correlation between liquidity and bid-ask spread and the positive correlation between liquidity and market depth.

### 6.2 Loss and Instantaneous Liquidity

For a market maker having a step-wise price function, Schwarz [2005] shows that the minimum worst-case loss for him to keep a bid-ask spread no greater than $s$ is bounded by $\frac{1}{8s}$ if he adjusts prices after every share of trading. Schwarz only considers two-outcome markets. In this part, we explore the relationship between instantaneous liquidity and market maker's worst-case loss for markets with continuous and differentiable price functions over $N$ outcomes.

We consider a class of market makers whose cost functions are symmetric and second-order differentiable. A second-order differentiable cost function ensures that price functions are differentiable. A symmetric cost function means that if a quantity vector $\vec{q}'$ contains a permutation of the elements of $\vec{q}$, $C(\vec{q}')$ equals $C(\vec{q})$. A symmetric cost function automatically gives symmetric price functions, meaning that if $\vec{q}'$ is achieved by switching $q_i$ and $q_j$ in $\vec{q}$, $p_i(\vec{q})$ equals $p_j(\vec{q}')$. This further implies that when all $q_i$'s are equal, all $p_i$'s equal to $\frac{1}{N}$. Assuming the market is started with $\vec{q}^0 = \vec{0}$, the initial market prices are uniform. Without loss of generality, we further assume that the market maker normalizes his cost function to satisfy $C(\vec{0}) = 0$.

Using properties given in Property 5, we can characterize the maximum loss of the market maker.

**Lemma 7.** *With a second-order differentiable and symmetric cost function $C(\vec{q})$ and corresponding price functions $p_i(\vec{q})$, a market maker's worst-case loss is bounded by*

$$L_{max} = \int_0^{+\infty} \left(1 - p_i(\vec{q}^i)\right) dq_i, \qquad (19)$$

*where $\vec{q}^i$ is a vector whose $i$-th element is $q_i$ and all others are zero. Choice of $i$ is arbitrary.*

Lemma 7 greatly simplifies the analysis of worst-case loss. It indicates that instead of considering the multivariate price function $p_i(\vec{q})$, we only need to consider the univariate price function $p_i(\vec{q}^i)$. If plotting the price function $p_i(\vec{q}^i)$ in a x-y plane as in Figure 2, the worst-case loss of a market maker equals the area formed by the y-axis, the horizontal line $p_i = 1$, and the curve of $p_i(\vec{q}^i)$ in the first quarter. Based on this result, we present our theorems on the relationship between instantaneous liquidity and market maker worst-case loss.

**Theorem 8.** *For all market makers who utilize a second-order differentiable and symmetric cost function over $N$ outcomes and have a fixed bound on worst-case loss, no market maker can exhibit uniformly greater or equal instantaneous liquidity than a different market maker for all $\vec{q}$.*

Theorem 8, which is proved by contradiction, states an impossibility result — it is impossible for a market maker to have greater instantaneous liquidity in all regimes given a fixed bound of worst-case loss. For a two-outcome market, Figure 2 plots the univariate price function of outcome 1, $p_1(\vec{q}^1)$, for two market makers who have the same bound of worst-case loss. It shows that neither of the price functions consistently has a smaller slope, i.e. greater instantaneous liquidity. Comparing the logarithmic utility market maker and the logarithmic MSR (negative exponential utility) market maker, we find that the logarithmic MSR market maker has greater instantaneous liquidity when prices are close to 0.5, while the logarithmic utility market maker exhibits greater instantaneous liquidity near extreme prices.

**Theorem 9.** *For all market makers who utilize a second-order differentiable and symmetric cost function over $N$ outcomes and maintain an instantaneous liquidity level no lower than $\rho$ for all $\vec{q}$, the minimum worst-case loss is $\frac{(N-1)^2 \rho}{2N^2}$.*

Theorem 9 indicates that among all market makers who keep the same lowest liquidity level $\rho$, the market maker's worst-case loss is bounded below by the loss of a linear price function $p_i(\vec{q}^{\,i})$ whose slope is $\frac{1}{\rho}$. Comparing with the bound, $\frac{1}{8s}$, derived by Schwarz [2005] for market makers who utilize step-wise price functions to maintain a bid-ask spread no greater than $s$, Theorem 9 gives the same bound by setting $N = 2$ and $\rho = \frac{1}{s}$. The optimal market maker described by Schwarz indeed is a discrete approximation of a continuous linear price function.

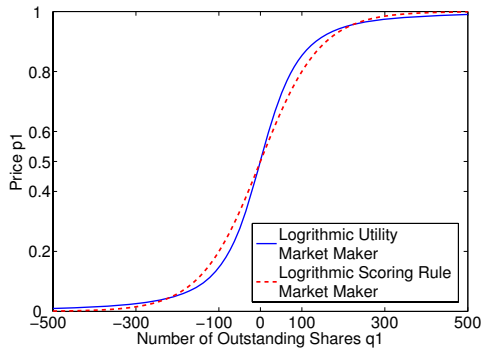

Figure 2: Price function $p_1(\vec{q}^{\,1})$ for a two-outcome market. Worst-case loss for both market makers are 50.

## 7 Conclusion

We introduce a class of utility-based market makers who are always willing to accept orders at their risk-neutral prices. We give the necessary and sufficient condition for such market makers to have bounded worst-case loss. We establish equivalence translations among utility-based market makers, market scoring rule market makers, and a third formulation of market makers based on maintaining a cost function. Specifically, (1) HARA utility market makers with parameter $\gamma$ are equivalent to weighted pseudospherical market scoring rule market makers with parameter $\beta = 1 - \frac{1}{\gamma}$ for $\gamma \neq 0$. (2) Translation between a utility-based market maker and the cost function formulation is based on the "expected utility keeps constant" equation, which implicitly defines the cost function; and (3) A market scoring rule market maker can be translated into a cost function formulation through a system of equations (16). We derive the corresponding cost and price functions for two special cases, logarithmic and quadratic scoring rules.

We show that for a fixed bound on worst-case loss, some market makers exhibit greater liquidity near uniform prices and some exhibit greater liquidity near extreme prices, but no market maker can exhibit uniformly greater liquidity in all regimes. For a fixed minimum liquidity level, we prove the lower bound for the worst-case loss of maker makers.

## 8 Acknowledgments

We thank Evdokia Nikolova.